\documentclass{article}
\usepackage[a4paper, portrait, margin=1.1811in]{geometry}
\usepackage[backend=biber, style=numeric,  sorting=none]{biblatex}
\usepackage[english]{babel}
\usepackage[utf8]{inputenc}
\usepackage[T1]{fontenc}
\usepackage{helvet}
\usepackage{etoolbox}
\usepackage{graphicx}
\usepackage{titlesec}
\usepackage{caption}
\usepackage{amsmath}
\usepackage{amssymb}
\usepackage{booktabs}
\usepackage{xcolor} 
\usepackage[colorlinks, citecolor=cyan]{hyperref}
\usepackage{caption}
\usepackage{subcaption}
\usepackage{authblk}
\usepackage{mdframed}
\graphicspath{ {./images/} }
\usepackage{scrextend}
\usepackage{fancyhdr}
\usepackage{graphicx}
\newcounter{lemma}

\newcounter{theorem}

\fancypagestyle{plain}{
	\fancyhf{}

	  \rhead{\small \textbf{DESY-23-107}}
}

\makeatletter
\patchcmd{\@maketitle}{\LARGE \@title}{\fontsize{16}{19.2}\selectfont\@title}{}{}
\makeatother

\newcommand{\dd}{\text{d}}
\newcommand{\av}[1]{\langle #1 \rangle}

\newcommand{\abs}[1]{\vert #1 \vert}

\newcommand{\firstdevp}[2]{\frac{\partial #1}{\partial #2}}
\newcommand{\seconddevp}[2]{\frac{\partial^2 #1}{\partial #2^2}}

\newcommand{\vb}[1]{\vec{#1}}
\newcommand{\cavvol}{V_{\text{cav}}}
\newcommand{\intcav}{\int_{V_{\text{cav}}}}
\newcommand{\intcavbound}{\int_{\partial V_{\text{cav}}}}

\newcommand{\cormode}[1]{\sigma#1^{(1)}_n}
\newcommand{\real}[1]{\text{Re}\big(#1\big)}

\newcommand{\cgmplus}{\eta^l_{+}}
\newcommand{\cgmtimes}{\eta^l_{\times}}
\newcommand{\cmem}{\eta^{l}_{01}}
\newcommand{\cgemefield}{\eta^{\text{E}}}
\newcommand{\cgembfield}{\eta^{\text{B}}}

\newsavebox\affbox
\author[1*]{\textbf{Robin Löwenberg}}
\author[1,2]{\textbf{Gudrid Moortgat-Pick}}
\affil[1]{ II. Institute for Theoretical Physics,
University of Hamburg, Luruper Chaussee 149, 22761 Hamburg, Germany
}
\affil[2]{ Deutsches Elektronen-Synchrotron DESY, Notkestr. 85, 22607 Hamburg, Germany
}

\titlespacing\section{0pt}{12pt plus 4pt minus 2pt}{0pt plus 2pt minus 2pt}
\titlespacing\subsection{12pt}{12pt plus 4pt minus 2pt}{0pt plus 2pt minus 2pt}
\titlespacing\subsubsection{12pt}{12pt plus 4pt minus 2pt}{0pt plus 2pt minus 2pt}

\titleformat{\section}{\normalfont\fontsize{10}{15}\bfseries}{\thesection.}{1em}{}
\titleformat{\subsection}{\normalfont\fontsize{10}{15}\bfseries}{\thesubsection.}{1em}{}
\titleformat{\subsubsection}{\normalfont\fontsize{10}{15}\bfseries}{\thesubsubsection.}{1em}{}

\titleformat{\author}{\normalfont\fontsize{10}{15}\bfseries}{\thesection}{1em}{}

\title{\textbf{\huge Lorentz Force Detuning in Heterodyne Gravitational Wave Experiments}}
\date{}    

\begin{document}

\pagestyle{headings}	
\newpage
\setcounter{page}{1}
\renewcommand{\thepage}{\arabic{page}}
	
\captionsetup[figure]{labelfont={bf},labelformat={default},labelsep=period,name={Figure }}	\captionsetup[table]{labelfont={bf},labelformat={default},labelsep=period,name={Table }}
\setlength{\parskip}{0.5em}
	
\maketitle
	
\noindent\rule{15cm}{0.5pt}
	\begin{abstract}
		Heterodyne cavity experiments for gravitational wave (GW) detection experience a rising interest since recent studies showed that they allow to probe the ultra high frequency regime above $10\,\text{kHz}$. In this paper, we present a concise theoretical study of the experiment based on ideas from the former MAGO collaboration which already started experiments in turn of the millenium. It extends the former results via deriving an additional term originating from a back-action of the electromagnetic field on the cavity walls, also known as Lorentz Force Detuning. We argue that this term leads to a complex dependence of the signal power $P_{\text{sig}}$ on the coupling coefficient between the mechanical shell modes and the electromagnetic eigenmodes of the cavity. It turns out that one has to adapt the coupling over the whole parameter space since the optimal value depends on the mechanical mode $\omega_l$ and the GW frequency $\omega_g$. This result is particularly relevant for the design of future experiments. 
  \\ \\
		\let\thefootnote\relax\footnotetext{
			\small $^{*}$\textbf{Corresponding author;} E-mail address: \textit{robin.loewenberg@desy.de} \\
		}
		\textbf{\textit{Keywords}}: \textit{Gravitational Waves; Lorentz Force Detuning; SRF Cavities; MAGO}
	\end{abstract}
\noindent\rule{15cm}{0.4pt}

\section{Introduction}
\begin{figure}[t]
    \centering
    \includegraphics[width=\textwidth]{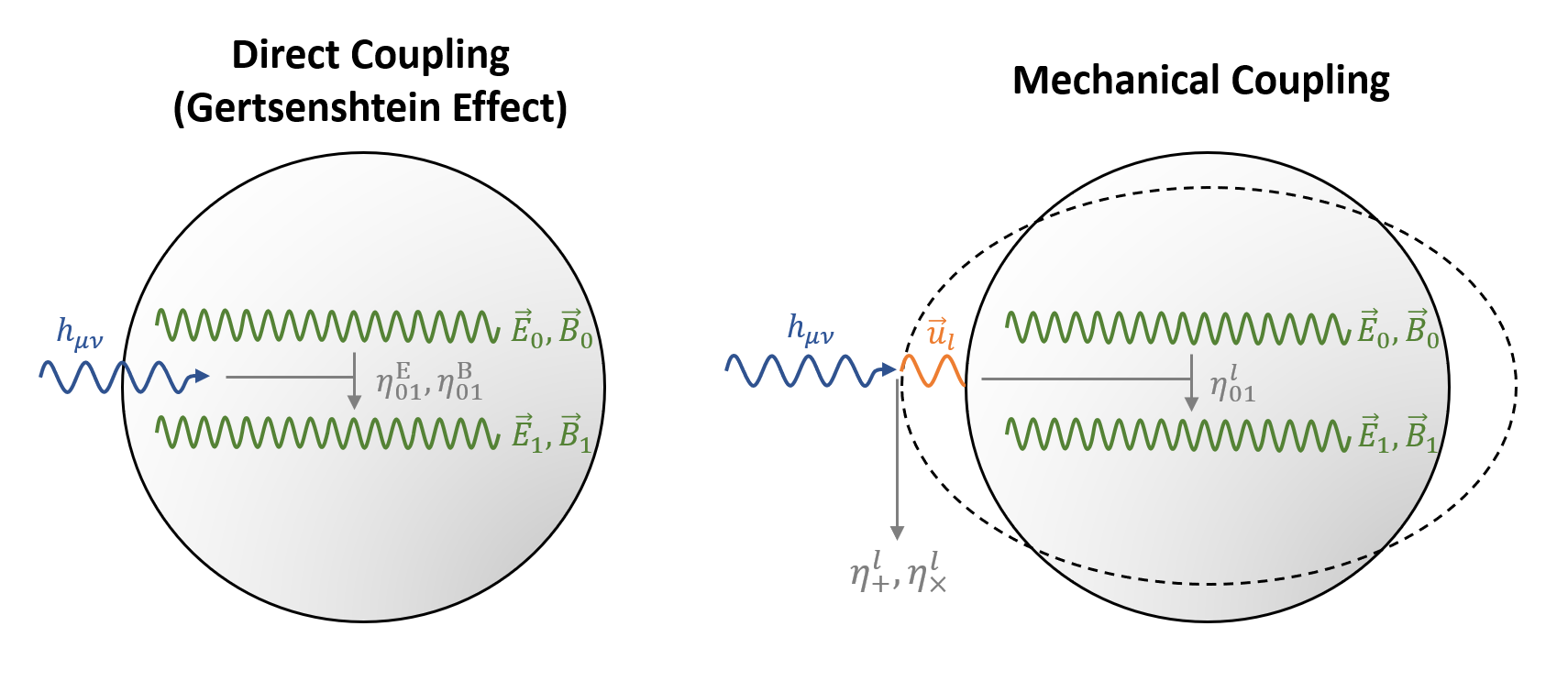}
    \caption{\textit{This sketch shows the two different channels of the GW interaction with the electromagnetic field. On the left, the direct coupling via the inverse Gertsenhtein effect is illustrated governed by two couplings $\cgembfield_{01}$ and $\cgemefield_{01}$ in the long wavelength regime. Note that $h_{\mu\nu}$ denotes the strain of the GW, which is supposed to be monochromatic throughout this study. On the right, the indirect coupling of the GW over a mechanical mode $\vb{u}_l$ is shown. It can be understood as a graviton-phonon-photon interaction in the particle picture. The GW-mechanical coupling can be described by two coupling coefficients $\cgmplus$ and $\cgmtimes$, while the mechanical-EM coupling is governed by $\cmem$. In both pictures, $\vb{E}_0$, $\vb{B}_0$ and $\vb{E}_1$, $\vb{B}_1$ are the fields of the pump and signal mode, respectively.}}
    \label{fig:GWCouplings}
\end{figure}

Since the first detection of gravitational waves (GWs) in 2016 by the LIGO and Virgo collaboration \cite{ABBOTT2016}, there is a rising interest on GW experiments probing the ultra high frequency regime beyond $10\,\text{kHz}$ \cite{AGG2021}. Because no source in this regime is known in the standard model of particle physics and cosmology, a detection would point towards new physics. Recent studies \cite{BERLIN2023} showed that heterodyne experiments using superconducting radio frequency (SRF) cavities are able to push the sensitivity towards a promising window for new sources. 
In particular, black hole superradiance would produce a very suitable signal for this approach \cite{BRIT2020}. The idea was initially worked out in the 1970s by several studies \cite{BRAG1971, BRAG1974, CAVES1979, BERN1978, PEGO1978, PEGO1980}. First experiments began in 1984 \cite{REECE1984} which led to further efforts by the MAGO collaboration at INFN in the late 1990s \cite{BALLA2005, BERN2001, BERNARD1998, BERN1998, BERN2002}. The concept is based on two eigenmodes of an electromagnetic cavity which are nearly degenerate. One eigenmode is excited by an external oscillator (pump mode) whereas the other (signal mode) is coupled to a readout system that measures the electromagnetic field power. When a GW passes by, it can induce a transition of photons from the pump mode into the signal mode, leading to an enhanced power loss at the readout. The signal reaches a maximum when the GW is resonant to the frequency difference between both levels. This allows using superconducting radio frequency (SRF) cavities to scan over a wide frequency range from $1\,\text{kHz}$ to several GHz. Measurements in the superconducting state of the cavity allow for very high electromagnetic quality factors ($Q\sim10^{10}$) \cite{JACK2006, BERLIN2020, BERLIN2021} which are mandatory to distinguish the levels at small frequency differences. One should note that the heterodyne approach is comparable to classical Webar Bar detectors \cite{WEB1960, AST1993, AST1997, MAU1996, VIN2006}, however, it was shown in \cite{BERLIN2023} that higher sensitivities are reached, in particular for GWs above $10\,\text{kHz}$. For more details, see e.g. \cite{TOBAR1995, BLAIR1995, TOBAR2000} and references therein. \\
For the coupling of a GW to the EM field of the cavity, there are two possible channels. One is a direct coupling via the Gertsenshtein effect \cite{GERTS1962, ZEL1973}, the other is an indirect mechanical coupling where the GW leads to a deformation of the cavity boundaries inducing an overlap between the initial eigenmodes. In previous studies, the direct coupling was neglected since it is much weaker than the mechanical coupling at low frequencies. However, it becomes dominant at high frequencies above $1\,\text{GHz}$, which was already investigated in detail in \cite{BERLIN2021} for static B-field setups used in Axion experiments \cite{KHAT2021, ZHONG2018}. Although our main interest focuses on the lower frequency regime between $1\,\text{kHz}-10\,\text{MHz}$, we add it for a complete picture of the coupling phenomena. For simplicity, we apply the long wavelength approximation which allows to describe the coupling by two distinct coupling constants $\cgembfield_{01}$ and $\cgemefield_{01}$ where 0 and 1 refer to the pump and signal mode respectively. The mechanical coupling via the l-th mechanical eigenmode consists of a mechanical-EM part with constant $\cmem$ and a GW-mechanical part. In the monochromatic case, the latter decomposes into two constants $\cgmplus$ and $\cgmtimes$. They correspond to the two possible polarisations of a GW. A sketch explaining the principle of the heterodyne approach is shown in fig. \ref{fig:GWCouplings}.\\
Although the theoretical details were already worked out by the MAGO collaboration \cite{ BALLA2005, BERN2002} as well as in recent studies \cite{BERLIN2023}, one goal of our research is to provide a concise and full description of the experiment from first principles. This allows us to point out an important difference between the old and new publications. In \cite{BERLIN2023}, the field back-action of the EM-modes to the mechanical modes was not taken into account. This effect is commonly known as Lorentz force detuning and can be made plausible because the EM field counteracts the external changes induced by the GW, which is comparable to Lenz's law. In particular, it leads to a signal damping which depends on $\cmem$ and becomes dominant in the sub-MHZ regime and close to the resonance $\omega_l=\omega_g$. Thus, optimising the coupling to $|\cmem|\sim 1$ as it was suggested in \cite{BERLIN2023} does in general not provide the strongest signal in the cavity. 
\newline
The paper is organised as follows: In chapter \ref{sec:GertsenshteinEffect}, we introduce our implementation of the Gertsenshtein effect using the long-wavelength approximation. In chapter \ref{sec:WallDeformation} and \ref{sec:TidalForceDensity}, we introduce concepts from elasticity theory to describe the mechanical coupling via wall deformation, i.e. the cavity boundaries. The change of the EM-modes due to the change of boundaries can be described with cavity perturbation theory, which is introduced in chapter \ref{sec:PerturbationTheory}. With these ingredients, we can derive the equations of motion in chapter \ref{sec:EquationsOfMotion} which are solved for a monochromatic GW in z-direction in chapter \ref{sec:SolutionsForMonoGW}. Finally, we provide a detailed analysis of the damping term in chapter \ref{sec:DampingTerm}. This study has grown out of a Master's thesis \cite{LOEW2023}, where more details can be found.

\section{The Gertsenshtein-Effect}
\label{sec:GertsenshteinEffect}
Gravitational waves (GW) are usually described in the framework of linearised theory of general relativity \cite{FORTINI1990}, \cite{HA2003}, \cite{MTW1973}. In this regime, the metric decomposes into a minkowskian part and a small strain tensor $h_{\mu\nu}$, i.e. $g_{\mu\nu}=\eta_{\mu\nu}+h_{\mu\nu}$. In particular, we assume that $\abs{h_{\mu\nu}}\ll 1$ and $\abs{\partial_\alpha h_{\mu\nu}}\ll 1$. Throughout this study, we use the convention $(-,+,+,+)$ for $\eta_{\mu\nu}$. \\
The coupling between a GW and the electromagnetic field is governed by the Einstein-Maxwell action given by \cite{BERLIN2021}
\begin{equation*}
    S_{\text{EM}}=\int\dd^4x\sqrt{-g}\big( -\frac{1}{4}g^{\mu\alpha}g^{\nu\beta}F_{\mu\nu}F_{\alpha\beta} - g^{\mu\nu}j_\mu A_\nu\big).
\end{equation*}
In vacuum, where $j_\mu=0$, this equation leads to a Lagrangian of the form
\begin{equation}
    \label{eqn:GertsenshteinLagrangian}
    \mathcal{L}=-\frac{1}{4}F_{\mu\nu}F^{\mu\nu}-j^{\mu}_{\text{eff}}A_\mu,
\end{equation}
where the effective current $j_{\text{eff}}$ is induced by the strain $h_{\mu\nu}$. Considering the explicit form \cite{BERLIN2021}
\begin{equation}
    \label{eqn:GertsenshteinCurrent}
    j^{\mu}_{\text{eff}}=\partial_\nu\big( \frac{{h^\alpha}_\alpha}{2}F^{\mu\nu}+{h^\nu}_\alpha F^{\alpha\mu}-{h^\mu}_\alpha F^{\alpha\nu}\big)
\end{equation}
of this current, we see that it does not transform covariantly like a four-vector. Hence, it is not invariant under coordinate transformations and we must therefore carefully think about the reference frame when evaluating the strain. It turns out \cite{BERLIN2021, RAKH2014} that the best choice is given by the proper detector frame. It is an example for a local Lorentz frame and therefore encodes the physical change of the detector due to a passing GW. Since we focus on GWs in the kHz-MHz regime, we apply the long-wavelength approximation which allows us to use the relatively simple metric
\begin{equation}
    \label{eqn:PDFMetric}
    \dd s^2=-\dd t^2\big( 1-\frac{1}{2}\ddot{h}^{\text{TT}}_{ij}(g)x^i x^j\big)+\dd x^i\dd x^j\delta_{ij},
\end{equation}
where $h^{TT}_{ij}$ is the strain in the Transverse Traceless (TT) and $g$ denotes the reference geodesic, i.e. the worldline of the detector. More details can be found in \cite{RAKH2014, CAP2020, MARZ1994, MAG2007} and in Appendix \ref{app:PDF}. \\
For simplicity, we will often refer to a monochromatic GW with frequency $\omega_g$ travelling in z-direction. In TT gauge and using complex notation, it can be written as 
\begin{equation}
    \label{eqn:StrainMonoGW}
    h^{TT}_{ij}(t)=\begin{pmatrix} h_+ & h_\times & 0 \\ h_\times & -h_+ & 0 \\ 0 & 0 & 0 \end{pmatrix}e^{i\omega_g t},
\end{equation}
where the spatial dependence are neglected in the long wavelength approximation. From eqn. \ref{eqn:PDFMetric} we can obtain the only non-vanishing component $h_{00}$ of the strain in the proper detector frame. It yields
\begin{equation}
    \label{eqn:MonochromaticGWZDirection}
    h_{00}(t,\vb{x})=-\frac{\omega^2_g}{2}(h_+(x^2-y^2)+2xyh_{\times})e^{i\omega_g t}=:-H_0(\vb{x})e^{i\omega_g t},
\end{equation}
where we have defined the function $H_0(\vb{x}):=\omega^2_g/2(h_+(x^2-y^2)+2xyh_{\times})$.\\
Finally, we can derive the equations of motion from eqn. \ref{eqn:GertsenshteinLagrangian} and \ref{eqn:GertsenshteinCurrent}. The resulting modified Maxwell equations have the form
\begin{equation}
    \label{eqn:MaxwellEinsteinEquations}
    \begin{array}{rl} \nabla\cdot\vb{E}&=\rho_{\text{eff}}, \\
    \nabla\times\vb{B}-\partial_t\vb{E}&=\vb{j}_{\text{eff}}, \end{array}
\end{equation}
with the effective charge density $\rho_{\text{eff}}$ and current density $\vb{j}_{\text{eff}}$. In the long wavelength regime, they yield
\begin{align}
    \label{eqn:EffCharge}
    \rho_{\text{eff}}&:=\frac{1}{2}\nabla(h_{00}), \\
    \label{eqn:EffCurrent}
    \vb{j}_{\text{eff}}&:=-\frac{1}{2}\partial_t(h_{00}\vb{E}_0)-\frac{1}{2}\nabla\times(h_{00}\vb{B}_0),
\end{align}
where $\vb{E}_0$ and $\vb{B}_0$ are the electromagnetic fields of the pump mode.

\section{Wall Deformation}
\label{sec:WallDeformation}
Since GWs change the spacetime metric, the cavity boundaries get modified. A GW couples to the mechanical modes of the resonator which shifts the EM eigenmodes inside the cavity. In order to describe this effect properly, we need to apply the framework of classical elasticity theory \cite{LALIF1987, THBLAN2021}. A concise formalism was already derived in \cite{BERN2002, LOBO1995} and we only give a short review here. \\
The starting point is the equation of motion for an isotropic elastic solid under the influence of an external force density $\vb{f}(t,\vb{x})$, i.e. 
\begin{equation}
    \label{eqn:EOMDeformationField}
     \rho(\vb{x})\seconddevp{\vb{u}(t,\vb{x})}{t} - (\lambda+\mu)\nabla(\nabla\vb{u}(t,\vb{x}))-\mu\nabla^2\vb{u}(t,\vb{x}) = \vb{f}(t,\vb{x}).
\end{equation}
Here, $\rho(\vb{x})$ denotes the material density and $\lambda$ and $\mu$ are the materials first and second Lam\'{e} parameters \cite{LOBO1995}. For the initial conditions $\vb{u}(\vb{x},0)=0=\firstdevp{\vb{u}}{t}(\vb{x},0)$, we can use the ansatz
\begin{equation}
    \label{eqn:Decomposition}
    \vb{u}(\vb{x},t)=\sum_{l=1}^{\infty}\vb{\xi}_l(\vb{x})q_l(t).
\end{equation}
This leads to a set of equations of motion for the $q_l(t)$ of the form 
\begin{equation}
    \label{eqn:IdealizedVibration}
    \ddot{q}_l(t)+\omega_l^2 q_k(t)=\frac{f_l(t)}{M},
\end{equation}
where M is the cavity mass and $f_l(t)$ the generalised force density. It is defined via the integral
\begin{equation}
    \label{eqn:GeneralizedForceDensity}
    f_l(t):=\intcav\dd^3x\vb{f}(t,\vb{x})\vb{\xi}_l(\vb{x})
\end{equation}
over the cavity volume $V_{\text{cav}}$. Finally, we note that the spatial modes $\vb{\xi}_l(\vb{x})$ are the eigensolutions of the equation
\begin{equation*}
    \label{eqn:VibrationEigenequation}
    \omega_l^2\rho(\vb{x})\vb{\xi}_l(\vb{x})+(\lambda+\mu)\nabla(\nabla\vb{\xi}_l(\vb{x})) +\mu\nabla^2\vb{\xi}_l(\vb{x})=0
\end{equation*}
and are therefore independent of $\vb{f}(t,\vb{x})$. They are normalised as
\begin{equation*}
    \intcav\dd^3x\vb{\xi}_k(\vb{x})\vb{\xi}_l(\vb{x})\rho(\vb{x})=M\delta_{kl}.
\end{equation*}
In general, those modes have to be determined numerically. An analytic solution for a spherical geometry can be found in \cite{LOBO1995}. 

\section{Tidal Force Density for Monochromatic Gravitational Waves}
\label{sec:TidalForceDensity}
Here, we want to state a simple example of the tidal force density $\vb{f}(t,\vb{x})$ induced by a passing GW. It can be derived from the equation of geodesic deviation \cite{MTW1973} and yields \cite{BERN2002, LOBO1995}
\begin{equation*}
    \vb{f}(t,\vb{x}) = -\rho(\vb{x}) R_{0i0j}(t)x_j\vb{e}_i,
\end{equation*}
where $\rho(\vb{x})$ is again the material density and $R_{0i0j}$ the Riemann curvature tensor. In case of a monochromatic GW in z-direction (eqn. \ref{eqn:StrainMonoGW}), we can further evaluate this expression and plug it into the generalised force density in eqn. \ref{eqn:GeneralizedForceDensity}. We get
\begin{equation}
    \label{eqn:GeneralizedForceDensiteZ}
    f_l(t)=-\frac{1}{2}\omega_g^2M V^{1/3}_{\text{cav}}\big(h_{+}\cgmplus+h_{\times}\cgmtimes\big)e^{i\omega_g t}=:F_l(t)e^{i\omega_g t}.
\end{equation}
The dimensionless coupling coefficients $\cgmplus$ and $\cgmtimes$ encode the coupling strength between the GW and the mechanical mode $l$. They are defined by
\begin{align*}
    \cgmplus&:=\frac{V^{-1/3}_{\text{cav}}}{M}\intcav\dd^3x\rho(\vb{x})\big( x\xi_{l,x}(\vb{x})-y\xi_{l,y}(\vb{x}) \big), \\
    \cgmtimes&:=\frac{V^{-1/3}_{\text{cav}}}{M}\intcav\dd^3x\rho(\vb{x})\big( x\xi_{l,y}(\vb{x})+y\xi_{l,x}(\vb{x}) \big).
\end{align*}

\section{Mode Decomposition and Cavity Perturbation Theory}
\label{sec:PerturbationTheory}
The electromagnetic field in an evacuated cavity is, without any perturbation, given by the wave equations
\begin{equation}
    \label{eqn:Eigensolutions}
    \Delta\vb{E}=\frac{1}{c^2}\seconddevp{\vb{E}}{t} \qquad \qquad \Delta\vb{B}=\frac{1}{c^2}\seconddevp{\vb{B}}{t}.
\end{equation}
with the boundary conditions \cite{JACK2006}
\begin{equation}
    \label{eqn:BoundaryConditions}
    \vb{n}\times\vb{E}|_S = 0, \qquad \vb{n}\cdot\vb{B}|_S=0,
\end{equation}
where $\vb{n}$ is the normal vector of the cavity shell $S$. The eigensolutions of this boundary value problem can be separated into a dimensionless time-dependent part $e_n(t)$, $b_n(t)$ and a spatial part $\vb{E}_n(\vb{x})$, $\vb{B}_n(\vb{x})$. The general solution can then be decomposed as
\begin{equation}
    \label{eqn:ModeExpansions}
    \vb{E}(t,\vb{x}) = \sum_n e_n(t)\vb{E}_n(\vb{x}), \qquad \qquad \vb{B}(t,\vb{x}) = \sum_n b_n(t)\vb{B}_n(\vb{x}).
\end{equation}
The normalisation of the spatial modes is chosen accordingly to \cite{BERLIN2019},
\begin{equation}
    \label{eqn:Normalization}
    \intcav\dd^3x\varepsilon_0\vb{E}_n\vb{E}_m = 2U_n\delta_{nm} = \intcav\dd^3x\frac{1}{\mu_0}\vb{B}_n\vb{B}_m,
\end{equation}
where $U_n$ is the average energy in mode $n$. Correspondingly, the time-dependent modes are given by
\begin{equation}
    \label{eqn:TimeExpansion}
    e_n(t)=\frac{\varepsilon_0}{2U_n}\intcav\dd^3x\vb{E}(t,\vb{x})\vb{E}_n(\vb{x}), \qquad \qquad b_n(t)=\frac{1}{2\mu_0 U_n}\intcav\dd^3x\vb{B}(t,\vb{x})\vb{B}_n(\vb{x}).
\end{equation}
When a GW is passing through the cavity, the geometry and therefore the boundary conditions for the electromagnetic field change. In particular, that means we have to change the set of boundary conditions. However, since the GW strain is very small ($\lesssim\mathcal{O}(10^{-21})$), cavity perturbation theory (CPT) can be applied. That means, the perturbed modes can be expanded in terms of the unperturbed modes. \\
It should be noted that there are some pitfalls when applying CPT to the spatial modes. The main problem is that such a series expansion would generally not fulfill the boundary conditions
\begin{equation*}
    \vb{n}\times\vb{E}(\vb{x})|_{S'} = 0 \qquad \vb{n}\cdot\vb{B}(\vb{x})|_{S'}=0.
\end{equation*}
Hence, in order to obtain a consistent theory, a perturbation theory is used only for time dependent modes. Applying CPT to the spatial modes in eqn. \ref{eqn:ModeExpansions} leads to wrong signs in the final result \cite{BERN2002}.
The CPT method we use has been derived in \cite{GOUBAU1961}. Further details can be found in Appendix \ref{app:CavityPerturbationTheory}. Note that we consider GWs with frequencies much smaller than the mode frequencies, so we can treat the shell displacement in adiabatic approximation. Applying the CPT formalism for the time-dependent modes $e_n(t)$ and $b_n(t)$ leads to
\begin{align}
    \label{eqn:EFieldExpansion}
    e'_n(t)&=e_n(t)+\sum_{m\neq n}\frac{U_m}{U_n}\alpha_{nm}e_m, \\
    \label{eqn:BFieldExpansion}
    b'_n(t)&=b_n(t)-\frac{1}{2}\mathcal{C}_{nn}b_n+\sum_{m\neq n}\frac{U_m}{U_n}\beta_{nm}b_m, \\
    \label{eqn:FrequencyExpansion}
    \omega'_n&=\omega_n-\frac{1}{2}\omega_n \mathcal{C}_{nn}.
\end{align}
The expansion coefficients are given by
\begin{equation*}
    \alpha_{nm}=\frac{\omega_n\omega_m}{\omega_m^2-\omega_n^2}\frac{U_m}{U_n}\mathcal{C}_{nm}, \qquad \qquad \beta_{nm}=\frac{\omega_n^2}{\omega_m^2-\omega_n^2}\frac{U_m}{U_n}\mathcal{C}_{nm},
\end{equation*}
where $\mathcal{C}_{nm}$ encodes all geometric and electromagnetic properties of the cavity. In a heterodyne setup, we can decompose this factor as
\begin{equation}
    \label{eqn:PureConnectionCoefficient}
    \mathcal{C}_{nm}=V^{-1/3}_{\text{cav}}\sqrt{\frac{U_n}{U_m}}\sum_l q_l(t)\eta^l_{nm},
\end{equation}
where the symmetric dimensionless coupling coefficient $\eta^l_{nm}$ can shown to be \cite{BERN2002}
\begin{equation}
    \label{eqn:ConnectionCoefficient}
    \eta^l_{nm}=\frac{V^{1/3}_{\text{cav}}}{2\sqrt{U_n U_m}}\intcavbound\dd\vb{S}\vb{\xi}_l(\vb{x})\Big[ \frac{1}{\mu_0}\vb{B}_n\vb{B}_m - \epsilon_0\vb{E}_n\vb{E}_m\Big].
\end{equation}

\section{The Equations of Motion}
\label{sec:EquationsOfMotion}
In order to derive the equations of motion (EoM), we use a similar formalism as in \cite{BERN2002}, however adding the direct coupling to the EM field via the Gertsenshtein effect. Although this coupling is subdominant at low frequencies, it becomes important at frequencies beyond $1\,\text{GHz}$. Since it could be necessary to extend the experimental search into this regime in the future, we include it already in this study. Our starting point is the extended Lagrangian 
\begin{equation}
    \label{eqn:FullLagrangian}
    L=\intcav\dd V\Big[ -\frac{1}{4}F'_{\mu\nu}F'^{\mu\nu}-\frac{1}{2}j^\mu_{\text{eff}}A'_\mu \Big]+\sum_l\big(\frac{1}{2}M\dot{q}_l^2(t)-\frac{1}{2}M\omega_lq^2_l(t)+q_l(t)f_l(t) \big),
\end{equation}
where the prime denotes the perturbed fields and $j^\mu_{\text{eff}}$ is given in eqn. \ref{eqn:EffCharge}-\ref{eqn:EffCurrent}. \\
We can now split the Lagrangian into two parts, $L= L_{\text{em}}+L_{\text{mech}}$, where $L_{\text{em}}$ describes dynamics of the EM-field and is given by
\begin{equation}
    \label{eqn:EMLagrangian}
    L_{\text{em}}=\intcav\dd V\Big[-\frac{1}{4} F'_{\mu\nu}F'^{\mu\nu}-\frac{1}{2}j_{\text{eff}}^{\mu}A'_{\mu}\Big].
\end{equation}
The Lagrangian $L_{\text{mech}}$ governs the physics of the mechanical displacement field and yields
\begin{equation}
    \label{eqn:MechLagrangian}
    L_{\text{mech}}=\sum_n 2U_n\big(e'^2_n(t)-b'^2_n(t)\big)+\sum_l\Big( \frac{1}{2}M\dot{q}_l^2(t)-\frac{1}{2}M\omega_l q_l^2(t)+q_l(t)f_l(t) \Big).
\end{equation}
Note that both $j^\mu_{\text{eff}}$ and the corrections to $A'_{\mu}$ are of order $\mathcal{O}(h)$, so we can drop the prime of the vector field in leading order and neglect the term here. With the techniques described in chapter \ref{sec:GertsenshteinEffect} to \ref{sec:PerturbationTheory}, it is straightforward to derive the equations of motion. For simplicity, we will assume that only one mechanical mode $l$ contributes to the dynamics throughout this study. Then, adding dissipative terms to account for the energy losses through the walls and by the external oscillator driving the pump mode, we find from eqn. \ref{eqn:EMLagrangian} that
\begin{align}
    \label{eqn:EOMPump}
    \ddot{b}_0+\frac{\omega_0}{Q_0}\dot{b}_0+\omega^2_0 b_0 &= \omega^2_0 V^{-1/3}_{\text{cav}}q_l\Big(\eta^l_{00}b_0+\sqrt{\frac{U_1}{U_0}}\cmem b_1 \Big) + J_0 + \frac{\omega_0}{Q_0}\sqrt{\frac{U_d}{U_0}}\dot{b}_{d}, \\
    \label{eqn:EOMSignal}
    \ddot{b}_1+\frac{\omega_1}{Q_1}\dot{b}_1+\omega^2_1 b_1 &= \omega^2_1 V^{-1/3}_{\text{cav}}q_l\Big( \eta^l_{11}b_1+\sqrt{\frac{U_0}{U_1}}\cmem b_0 \Big) 
    +J_1+\epsilon\frac{\omega_1}{Q_1}\sqrt{\frac{U_d}{U_1}}\dot{b}_d,
\end{align}
where $n=0$ refers to the pump mode and $n=1$ to the signal mode. Here, $Q_0$ and $Q_1$ are the quality factors \cite{JACK2006} of the eigenmodes and $b_d$ denotes the oscillator which is also coupled to the signal mode with a constant\footnote{For the MAGO cavity, this coupling could be reduced to $\epsilon\sim10^{-7}$.} $\epsilon$. The Gertsenshtein current shows up as a projected current $J_n$ which can be expressed as
\begin{equation}
    \label{eqn:MonoProjectedCurrent}
    J_n(\omega):= H\omega_g^2\sqrt{\frac{U_0}{U_n}}\big( \kappa_n\cgemefield_{0n} + \lambda_n\cgembfield_{0n} \big)2\pi\delta(\omega-(\omega_0+\omega_g))
\end{equation}
under the assumption that $\vb{n}\times\vb{j}_{\text{eff}}(t,\vb{x})\vert_S=0$. Here we have introduced another two coupling coefficients
\begin{align}
    \eta^{\text{E}}_{0n}&:=\frac{1}{H\sqrt{U_0 U_n}}\intcav\dd^3x H_0(\vb{x})\varepsilon_0\vb{E}_0(\vb{x})\vb{E}_n(\vb{x}) \\
    \eta^{\text{B}}_{0n}&:=\frac{1}{H\sqrt{U_0 U_n}}\intcav\dd^3x H_0(\vb{x})\frac{1}{\mu_0}\vb{B}_0(\vb{x})\vb{B}_n(\vb{x}).
\end{align}
for the E-field and the  B-field respectively. The parameters $\kappa_n$ and $\lambda_n$ are given by
\begin{align}
    \label{eqn:Kappa}
    \kappa_n &:= i\frac{\omega_n}{8c^2}(\omega_0+\omega_g), \\
    \label{eqn:Lambda}
    \lambda_n &:= \frac{\omega_n^2}{8c^2}.
\end{align}
Note that they are different in the long wavelength regime. If $\omega_0+\omega_g<\omega_n$, the GW couples stronger to the B-field than to the E-field, and vice versa for $\omega_0+\omega_g>\omega_n$. Since we work with $\omega_0\approx\omega_1$ and $\omega_g\ll\omega_0,\omega_1$ throughout this study, we assume $|\kappa_n|\approx|\lambda_n|$, i.e. that the couplings are equally strong. If $\omega_g$ becomes comparable to $\omega_0$ and $\omega_1$, the long wavelength approximation may break down and eqn. \ref{eqn:MonoProjectedCurrent} is no longer valid. Finally, we have defined the normalised GW strain as
\begin{equation}
    \label{eqn:StrainNorm}
    H:=\sqrt{\frac{1}{V_{\text{cav}}}\intcav\dd^3x H^2_0(\vb{x})}.
\end{equation}
For a MAGO-like cavity, we found that a reasonable value for the normalised strain is given by $H\sim h_0\times 10\,\text{m}^2$, where $h_0$ is the characteristic strain strength of the GW. \\
Similarly, we can find the EoM for the mechanical modes from eqn. \ref{eqn:MechLagrangian}. Adding again a dissipative term, we end up with
\begin{equation}
    \label{eqn:EOMQl}
    \ddot{q}_l+\frac{\omega_l}{Q_l}\dot{q}_l+\omega_l^2 q_l = \frac{1}{M}\Big(f_l+f_l^{\text{ba}}\Big),
\end{equation}
where $Q_l$ is the mechanical quality factor. In contrast to \cite{BERLIN2023}, we automatically obtain the field back-action 
\begin{equation*}
    f_l^{\text{ab}}(t):=V^{-1/3}_{\text{cav}}\Big(U_0 \eta^l_{00}b_0^2(t)+U_1\eta^l_{11}b_1^2(t)+2\sqrt{U_0 U_1}\cmem b_0(t)b_1(t)\Big),
\end{equation*}
which is responsible for an additional deformation of the cavity walls, commonly known as Lorentz Force Detuning. The first two terms lead to a constant shift that can be absorbed into the definition of the eigenmodes. However, the last term leads to a damping of the signal strength which is particularly strong close to the resonances. Effectively, we therefore get
\begin{equation}
    \label{eqn:BackAction}
    f_l^{\text{ab}}(t)=2V^{-1/3}_{\text{cav}}\sqrt{U_0 U_1}\cmem b_0(t)b_1(t).
\end{equation}
Note that this term already appeared in the studies of the MAGO collaboration, see e.g. \cite{BALLA2005, BERN2002}. However, in chapter \ref{sec:DampingTerm}, we will investigate its impact on the detector sensitivity in greater detail.

\section{Solution for Monochromatic Gravitational Waves}
\label{sec:SolutionsForMonoGW}
In general, the coupled differential equations \ref{eqn:EOMPump}, \ref{eqn:EOMSignal} and \ref{eqn:EOMQl} can only be solved by numerical methods. However, if we assume a monochromatic GW travelling in z-direction, i.e.
\begin{align*}
    q_l(t)&=\real{Q_l(t)e^{i\omega_gt}}, \\
    f_l(t)&=\real{F_l(t)e^{i\omega_gt}}, \\
    J_1(t)&=\real{K_1(t)e^{i(\omega_0+\omega_g)t}},
\end{align*}
it is possible to find an analytic expression for the signal power. We further assume that the pump mode can be stabilised to be monochromatic. Therefore, an appropriate ansatz for the fields is
\begin{align}
    \label{eqn:bFieldPump}
    b_0(t)=b_d(t)&=\real{e^{i\omega_0t}}, \\
    \label{eqn:bFieldSignal}
    b_1(t)&=\real{A_1(t)e^{i(\omega_0+\omega_g)t}}.
\end{align}
The remaining calculation is greatly inspired by \cite{BERN2002}. If $Q_l(t)$ and $A_1(t)$ are small such that only leading terms in these functions are relevant, the EoM (eqn. \ref{eqn:EOMSignal} and eqn. \ref{eqn:EOMQl}) can be written as
\begin{align}
    \label{eqn:TimeEquationForA1}
    \ddot{A}_1(t)+\alpha_1\dot{A}_1(t)+\beta_1A_1(t) &=\sqrt{\frac{U_0}{U_1}}\gamma_1Q_l(t)+K_1(t)+\epsilon i\omega_0\frac{\omega_1}{Q_1}\sqrt{\frac{U_d}{U_1}}e^{-i\omega_gt}, \\
    \label{eqn:TimeEquationForQl}
    \ddot{Q}_l(t)+\alpha_l\dot{Q}_l(t)+\beta_l\omega_l^2 Q(t) &= \frac{F_l(t)}{M} + \sqrt{\frac{U_1}{U_0}}\gamma_l A_1(t),
\end{align}
where we also neglected all fast oscillating terms. That means, we assumed that terms containing $e^{i\omega_0 t}$ or $e^{i\omega_1 t}$ vanish in the time average compared to terms containing $e^{i\omega_g t}$. The constants introduced in eqn. \ref{eqn:TimeEquationForA1} and \ref{eqn:TimeEquationForQl} are given by
\begin{align*}
    \alpha_l&:=2i\omega_g+\frac{\omega_l}{Q_l}, \\
    \beta_l&:=\omega_l^2-\omega_g^2+i\omega_g\frac{\omega_l}{Q_l}, \\
    \gamma_l&=\frac{1}{M}\cavvol^{-1/3}U_0\cmem, \\
    \alpha_1&:=\frac{\omega_1}{Q_1}+2i(\omega_0+\omega_g), \\
    \beta_1&:=\omega_1^2-(\omega_0+\omega_g)^2+i\frac{\omega_1}{Q_1}(\omega_0+\omega_g), \\
    \gamma_1&:=\cavvol^{-1/3}\omega^2_1\cmem.
\end{align*}
We can now perform Fourier transformations to solve eqn. \ref{eqn:TimeEquationForA1} and \ref{eqn:TimeEquationForQl} for $A_1(t)$. By again neglecting fast oscillating terms, the time average of $b_1(t)$, eqn. \ref{eqn:bFieldSignal}, can be expressed as $\av{b^2_1(t)}=\frac{1}{2}\av{|A_1(t)|^2}$. Thus, we can write the signal in terms of a power spectral density (PSD) 
\begin{equation*}
    S_{\text{sig}}(\omega):=2U_1\omega_1/Q_{\text{cpl}} S_{b_1}(\omega),
\end{equation*}
where $\av{b^2_1(t)}=(2\pi)^{-2}\int\dd\omega S_{b_1}(\omega)$. Note the coupling quality factor $Q_{\text{cpl}}$ has to be used here, since it parameterises the energy transfer into the readout system. It is related to the full quality factor $Q_1$ via
\begin{equation*}
    \frac{1}{Q_1}=\frac{1}{Q_{\text{cpl}}}+\frac{1}{Q_{\text{int}}},
\end{equation*}
where $Q_{\text{int}}$ is the internal quality factor neglecting the readout loss. More details on that can be found in \cite{BERLIN2023, BERLIN2020}. \\
The final result for the signal PSD is then given by 
\begin{equation}
    \label{eqn:SignalPSD}
    S_{\text{sig}}(\omega)=\frac{\omega_1}{Q_{\text{cpl}}}\omega_g^4U_0\Bigg| \underbrace{\frac{1}{2}\frac{\omega_1^2\cmem(h_+\cgmplus + h_{\times}\cgmtimes)}{\Lambda_{1}(\omega-(\omega_0+\omega_g))}}_{\text{Mechanical Coupling}}-\underbrace{\frac{H(\kappa_1\cgemefield_{01}+\lambda_1\cgembfield_{01})}{\Lambda_2(\omega-(\omega_0+\omega_g))}}_{\text{Gertsenshtein Coupling}}\Bigg|^24\pi^2\delta(\omega-(\omega_0+\omega_g)).
\end{equation}
where we have introduced two functions $\Lambda_1(\omega)$ and $\Lambda_2(\omega)$ in the denominators which we call \textit{resonance functions}. They are given by
\begin{align}
    \label{eqn:Lambda1}
    \Lambda_1(\omega)&:=\big( \beta_1-\omega^2+i\omega\alpha_1 \big)\big( \beta_l-\omega^2+i\omega\alpha_l \big)-\gamma_1\gamma_l, \\
    \label{eqn:Lambda2}
    \Lambda_2(\omega)&:=\Lambda_1(\omega)\big( \beta_l-\omega^2+i\omega\alpha_l \big)^{-1}.
\end{align}
For completeness, we note that an additional term appears from eqn. \ref{eqn:TimeEquationForA1}, describing the coupling to the external oscillator. It yields
\begin{equation}
    \label{eqn:OscillatorPSD}
    S_{\text{osc}}(\omega)=\epsilon^2 \frac{Q_1}{Q_{\text{cpl}}}\frac{\omega^3_1}{Q^3_1}\omega_0^2\frac{U_d S_{b_d}(\omega)}{|\Lambda_2(\omega-(\omega_0+\omega_g))|^2},
\end{equation}
where we have defined $S_{b_d}(\omega):=4\pi^2\delta(\omega-\omega_0)$. For a monochromatic oscillator, this PSD can be well separated from the signal. However, there is some irreducible phase noise which leads to a power leakage into frequency range of the signal mode. It turns out that, for $\epsilon\sim10^{-7}$, the oscillator phase noise is negligible compared to more dominant sources such as mechanical or thermal noise. For a detailed discussion, see e.g. \cite{BERLIN2023, BERLIN2019}. \\
Finally, we can integrate eqn. \ref{eqn:SignalPSD} to obtain the total signal power. It yields
\begin{equation}
    \label{eqn:SignalPower}
    P_{\text{sig}}=\frac{1}{(2\pi)^2}\int\dd\omega S_{\text{sig}}(\omega) = \frac{\omega_1}{Q_{\text{cpl}}}\omega_g^4U_0\Bigg| \frac{1}{2}\frac{\omega_1^2\cmem(h_+\cgmplus + h_{\times}\cgmtimes)}{\beta_1\beta_l-\gamma_1\gamma_l}-\frac{\beta_lH(\kappa_1\cgemefield_{01}+\lambda_1\cgembfield_{01})}{\beta_1\beta_l-\gamma_1\gamma_l} \Bigg|^2.
\end{equation}
This is the main result of our study. In contrast to the results of \cite{BERLIN2023}, it shows an additional damping factor
\begin{equation}
    \label{eqn:DampingTerm}
    \gamma_1\gamma_l=\frac{1}{M}\cavvol^{-2/3}U_0(\omega_1\cmem)^2
\end{equation}
in the denominator. In the following, we will investigate its impact on the detector sensitivity.

\section{Impact of the Damping Term}
\label{sec:DampingTerm}
\begin{figure}[b!]
    \centering
    \begin{subfigure}[b!]{0.49\textwidth}
        \centering
        \includegraphics[width=\textwidth]{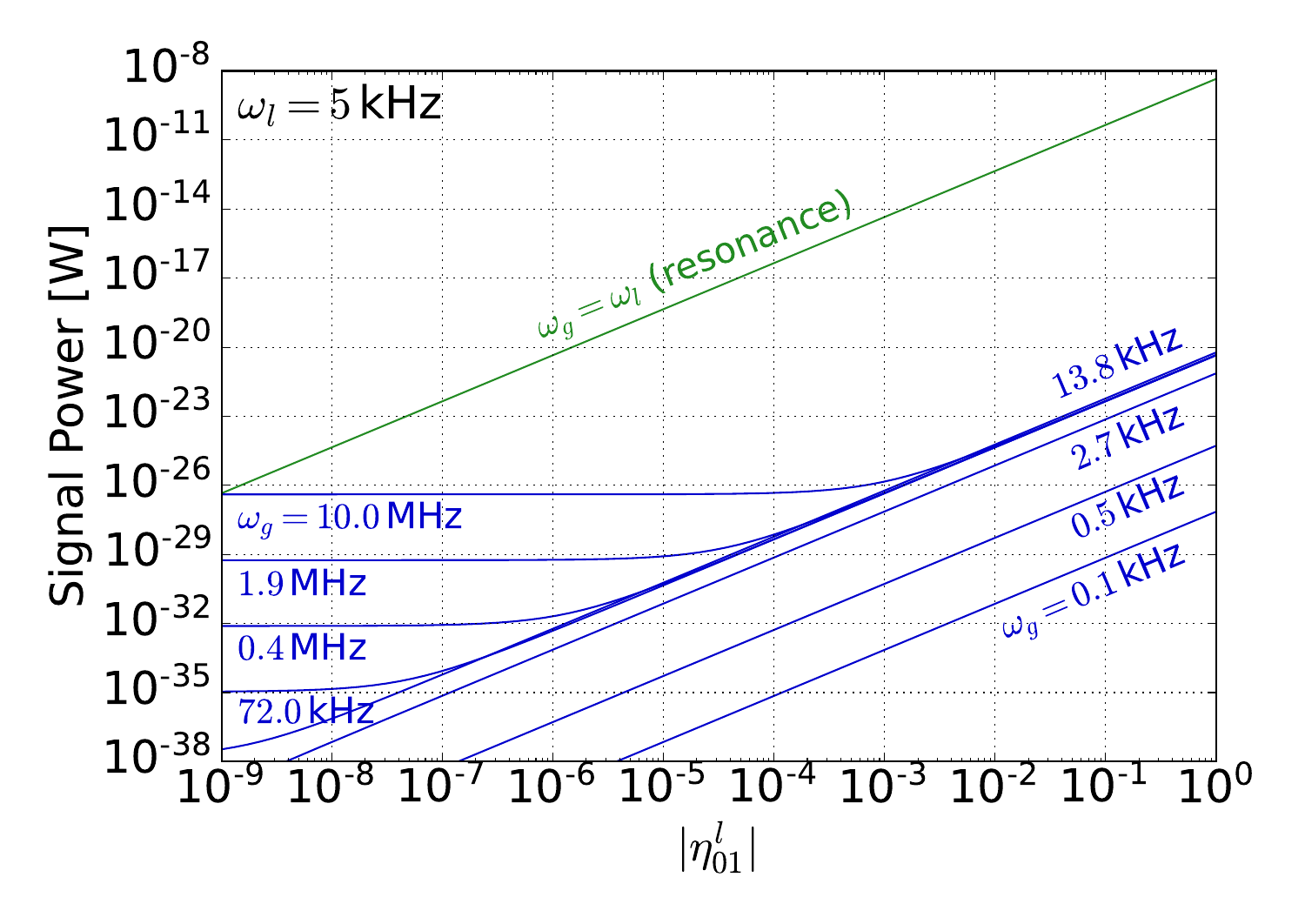}
        \caption{\textit{Signal without damping}}
    \end{subfigure}
    \begin{subfigure}[b!]{0.49\textwidth}
        \centering
        \includegraphics[width=\textwidth]{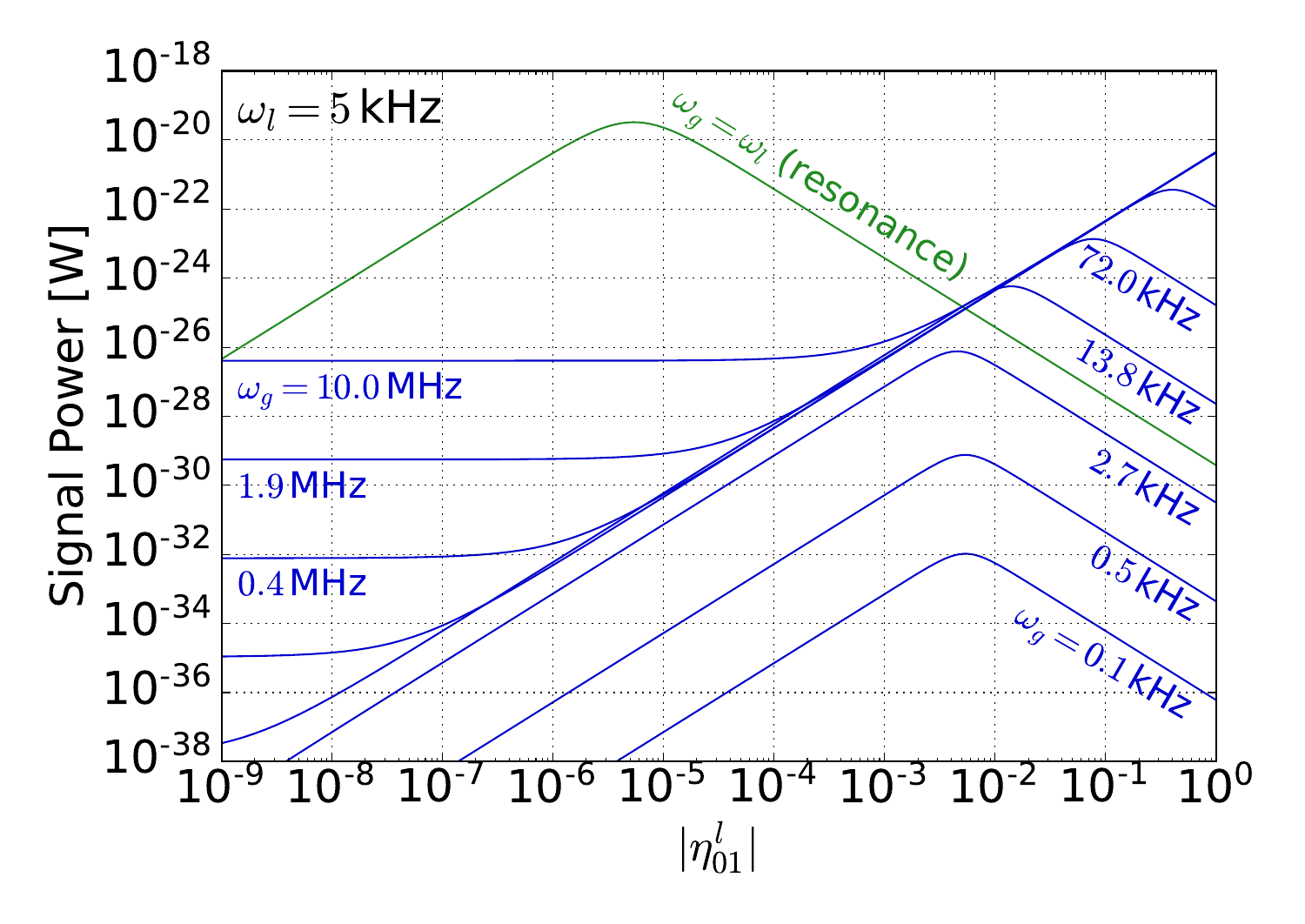}
        \caption{\textit{Signal with damping}}
    \end{subfigure}
    \caption{\textit{This plot displays how the signal power depends on the mechanical-em coupling $\cmem$. The curves were calculated for different values of $\omega_g$ between $100\,\text{Hz}$ and $10\,\text{MHz}$ assuming a scanning experiment (i.e. $\omega_1=\omega_0+\omega_g$).  \textbf{(a)}: Calculations without the damping term. As expected, the highest possible signal is always achieved for $|\cmem|=1$. \textbf{(b)}: Calculations including the damping term, eqn. \ref{eqn:DampingTerm}. A remarkable result is that, in particular for $\omega_g$ below $\sim1\text{\,\text{MHz}}$ and close to the resonance $\omega_g=\omega_l$ (green curve), the best signal power is achieved for $|\cmem|<1$.}}
    \label{fig:SignalCouplingRelation}
\end{figure}
\begin{figure}[t!]
    \begin{center}
        \includegraphics[width=0.7\textwidth]{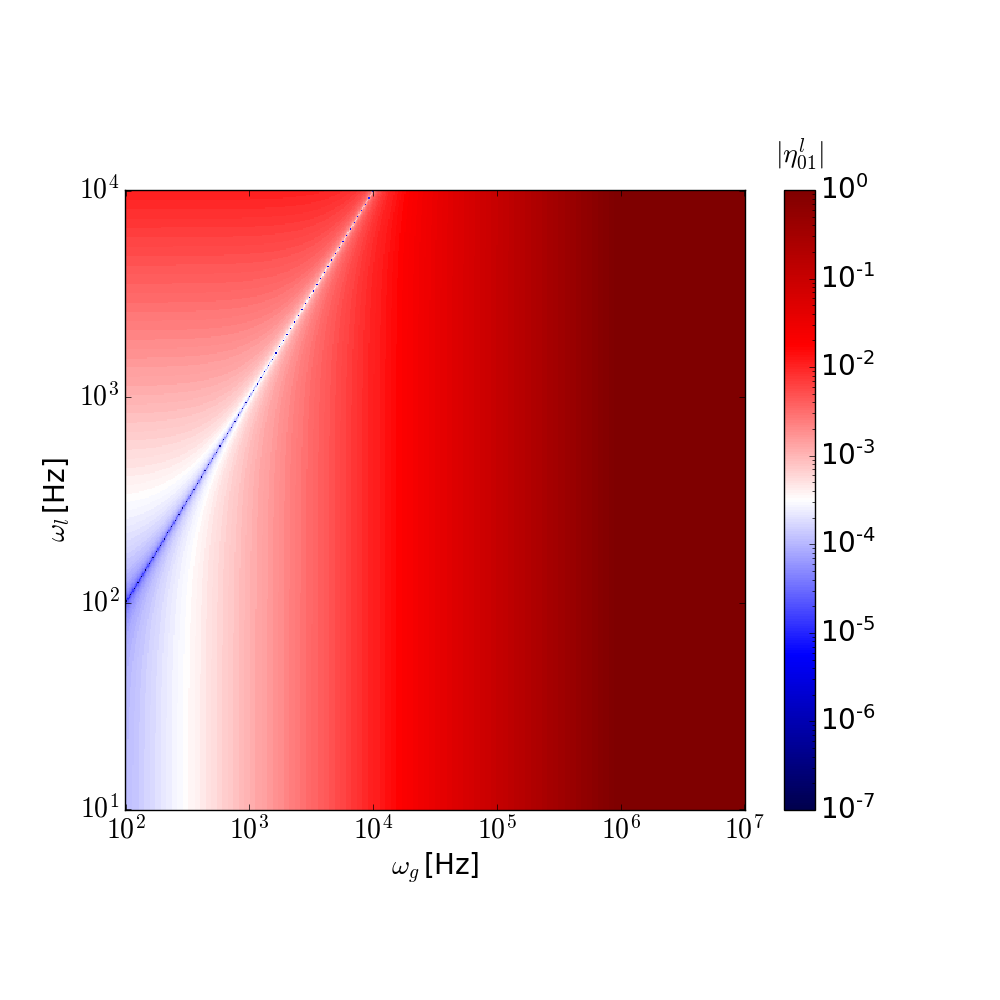}
        \caption{\textit{This plot shows the coupling strength leading to the highest signal power in the $\omega_g$-$\omega_l$-plane. Some examples for fixed $\omega_g$ and $\omega_l$ are shown in fig. \ref{fig:SignalCouplingRelation}. An important result of this plot is that the highest signal is not always reached for $|\cmem|=1$ as proposed by \cite{BERLIN2023}. For $\omega_g$ in the sub-MHz regime, the optimal coupling can be much lower. In particular close to the resonance $\omega_g=\omega_l$, the coupling should be optimised to values of order $|\cmem|\sim\mathcal{O}(10^{-6})$.}}
        \label{fig:OptimisationCurve}
    \end{center}
\end{figure}

In this section we want to investigate the effects of the damping term $\gamma_1\gamma_l$ which follows from the back-action $f^{\text{ab}}_l(t)$, eqn. \ref{eqn:BackAction}. The main reason is that recent studies, such as \cite{BERLIN2023}, do not consider this term, although it has an important influence on the results, as explained in the following. The MAGO collaboration, however, mentioned it, but without investigating it in greater detail. \\
We found that the term has a great influence on the signal since it depends quadratically on the mechanical coupling. Therefore, we propose that a mechanical coupling of $|\cmem|=1$, cf. eqn. \ref{eqn:ConnectionCoefficient}, does not always lead to the strongest signal. To show this, we calculate eqn. \ref{eqn:SignalPower} explicitly and take $\omega_l$, $\omega_g$ and $\cmem$ as free parameters. In order to fix the remaining values, we mostly follow \cite{BERLIN2023, BALLA2005}. That means, concerning the cavity parameters, we choose $M=5\,\text{kg}$, $\cavvol=10\,\text{L}$, $\omega_0=1.8\,\text{GHz}$ and $Q_0=Q_{\text{cpl}}=Q_{\text{int}}=10^{10}$. The electromagnetic field in the cavity and the temperature of the boundaries should not exceed the quenching limit of niobium, which was used for MAGO. According to \cite{BERLIN2023}, we therefore assume a typical E-field of $30\,\text{MV/m}$ which corresponds to a total pump mode energy of $U_0\sim 40\,\text{J}$. For the GW, we assume a typical strain of $h_0=h_+=h_{\times}=10^{-20}$ and a GW-mechanical coupling of $\cgmplus=\cgmtimes=1$, which are both rather optimistic\footnote{For MAGO-like cavities, we found values of $\cgmplus,\cgmtimes\sim\mathcal{O}(10^{-2})$. Note, however, that our qualitative results do not depend on these values.}. The calculations are conducted for a scanning experiment where $\omega_1=\omega_0+\omega_g$. Broadband detection is in principle possible as well \cite{BERLIN2023, BERLIN2020}, but typically leads to a low sensitivity far from the resonance. \\
In the first analysis, we assume a lowest mechanical quadrupole mode at $\omega_l=5\,\text{kHz}$ which is in good agreement with numerical simulations of MAGO-like cavity spectra. We then investigated how the signal power depends on $|\cmem|$ for eight different frequencies $\omega_g$ between $100\,\text{Hz}$ and $10\,\text{MHz}$. For larger frequencies, the long wavelength approximation may break down and eqn. \ref{eqn:SignalPower} has to be adjusted. The results are shown in fig. \ref{fig:SignalCouplingRelation} both with and without the damping term. When the term is neglected (fig. \ref{fig:SignalCouplingRelation}, left panel), it is obvious that the highest signal is achieved for the maximum coupling constant $|\cmem|=1$. However, in case it is included (fig. \ref{fig:SignalCouplingRelation}, right panel), $|\cmem|=1$ leads to the strongest signal only for high frequencies in the MHz-regime. Below, we find that the best coupling is achieved for $|\cmem|<1$. In particular close to the resonance $\omega_g=\omega_l$, the damping term leads to an ideal coupling of $|\cmem|\sim\mathcal{O}(10^{-6})$. The results therefore clearly show that SRF cavity experiments should in general not be optimised to $|\cmem|\sim\mathcal{O}(1)$. This is particularly important for low frequencies and frequencies close to the resonance. \\
We also provide a more general analysis of the best choice for $|\cmem|$ in fig. \ref{fig:OptimisationCurve}. It shows the value of $|\cmem|$ with the largest signal powers in the $\omega_g$-$\omega_l$-plane. The result could be used as a template for optimising future gravitational wave experiments. We point out that the parameters $\omega_l$, $\omega_g$ and $\cmem$ can be controlled via the cavity geometry.
Finally, we note that the optimal coupling does also depend on the noise sources which determine the cavity sensitivity. Applying a qualitative analysis of the sources discussed in \cite{BERLIN2023} using the damping term showed that the coupling still has to be optimised similar to fig. \ref{fig:OptimisationCurve}. However, the results largely depend on the noise parameters and we postpone a more detailed study to future work.

\section{Conclusion}
\label{sec:Conclusions}
Heterodyne cavity experiments provide a promising tool for detecting or excluding new sources of GWs in the future. Recent studies showed that the sensitivity of modern cavities can already approach the regime for new physics. A promising candidate in the measurable regime is, for instance, black hole superradiance. In this study, we refined the theoretical formalism proposed by the former MAGO collaboration \cite{BALLA2005, BERN2002} and added another signal source from the Gertsenshtein effect. The resulting signal power for a monochromatic GW in z-direction is shown in eqn. \ref{eqn:SignalPower}. We note that the Gertsenshtein effect is subdominant in the considered kHz-MHz-regime. At higher frequencies above MHz, however, the effect becomes dominant and the coupling has to be taken into account. In that case, the long wavelength approximation breaks down and the full metric expansion should be used (see appendix \ref{app:PDF}). First studies can be found in \cite{BERLIN2023, BERLIN2021}. \\
An important difference to the recent results \cite{BERLIN2023} is that we included the back-action of the EM field, leading to Lorentz Force Detuning and causing a damping term $\gamma_1\gamma_l$, eqn. \ref{eqn:DampingTerm}, which depends on the coupling $\cmem$ of the EM field to the mechanical cavity modes. This term was already described in \cite{BALLA2005, BERN2002}, but its influence was not investigated further. We found that an important consequence is that choosing $|\cmem|\sim\mathcal{O}(1)$ does in general not lead to the strongest signal power. In particular close to the resonance $\omega_g=\omega_l$, the maximum signal is achieved for much lower couplings of order $|\cmem|\sim\mathcal{O}(10^{-6})$. \\
Altogether, we recommend that future heterodyne cavity experiments do not choose a design where always $|\cmem|\sim 1$. Instead, the coupling constant should be adjusted such that it matches the optimal coupling for the values $\omega_l$ and $\omega_g$ of the experiment.

\section{Acknowledgements}
We gratefully acknowledge helpful discussions with Marc Wenskat, Krisztian Peters, Andreas Ringwald, Lars Fischer, Michel Paulsen, Sebastian Ellis, Raffaele Tito D’Agnolo and Jan Schütte-Engel. GMP is supported by the Deutsche Forschungsgemeinschaft (DFG, German Research Foundation)
under Germany’s Excellence Strategy EXC 2121 ``Quantum Universe'' - 390833306.

\printbibliography

\newpage
\appendix
\section{Appendix A: The Proper Detector Frame}
\label{app:PDF}
The proper detector frame (PDF) is a coordinate system that describes the local Lorentz frame of an observer in curved spacetime \cite{MARZ1994}. With respect to the cavity, it properly describes the response of the EM-field to a small metric perturbation in its local lorentz frame. We therefore need it to describe the gauge-dependent Gertsenshtein current of a passing GW. \\
In general, the PDF is a combination of Fermi Normal Coordinates (FNC) \cite{MTW1973, MANMIS1963} and the acceleration and rotation of the observer \cite{NIZIM1978}. As shown by Marzlin in 1994 \cite{MARZ1994}, the metric can be then written as
\begin{align}
    g_{00}&=-(1+\vb{a}\cdot\vb{x})^2+(\vb{\omega}\times\vb{x})^2-\gamma_{00}-2(\vb{\omega}\times\vb{x})^i\gamma_{0i}-(\vb{\omega}\times\vb{x})^i(\vb{\omega}\times\vb{x})^j\gamma_{ij}\notag, \\
    \label{eqn:GeneralFermiNormalCoords}
    g_{0i}&=(\vb{\omega}\times\vb{x})_i-\gamma_{0i}-(\vb{\omega}\times\vb{x})^j\gamma_{ij}, \\
    g_{ij}&=\delta_{ij}-\gamma_{ij},\notag
\end{align}
where the coefficients are given by the series expansions
\begin{align*}
    \gamma_{00}&=\sum_{n=0}^\infty\frac{2}{(n+3)!}x^kx^lx^{k_1}\cdots x^{k_n}(\partial_{k_1}\cdots\partial_{k_n}R_{0k0l})(g)\cdot\Big[ (n+3)+2(n+2)\vb{a}\vb{x}+(n+1)(\vb{a}\vb{x})^2 \Big], \\
    \gamma_{0i}&=\sum_{n=0}^\infty\frac{2}{(n+3)!}x^kx^lx^{k_1}\cdots x^{k_n}(\partial_{k_1}\cdots\partial_{k_n}R_{0kil})(g)\cdot\Big[(n+2)+(n+1)\vb{a}\vb{x}\Big], \\
    \gamma_{ij}&=\sum_{n=0}^\infty\frac{2}{(n+3)!}x^kx^lx^{k_1}\cdots x^{k_n}(\partial_{k_1}\cdots\partial_{k_n}R_{ikjl})(g)\cdot\Big[ n+1 \Big],
\end{align*}
where $\vb{a}$ and $\vb{\omega}$ are the acceleration and rotation and $g$ the geodesics of the observer (reference geodesic). 
Note that the Riemann tensor is gauge independent. For GWs, it is therefore possible to compute it in the more convenient TT-gauge. \\
With SRF experiments, the goal is to measure GWs with frequencies 
in the range $\sim\mathcal{O}(\text{kHz-MHz})$, which is much above the typical variations of the gravitational field on earth with values of $f\lesssim 0.1\,\text{Hz}$ \cite{CAP2020}. We can therefore well separate the GWs from the background field and set $\vb{a}=0$ and $\vb{\omega}=0$. The resulting simplified expansion for the GW strain reads
\begin{align}
    \label{eqn:FullResult00Component}
    h_{00}&=-2\sum_{n=0}^\infty\frac{n+3}{(n+3)!}x^kx^lx^{k_1}\cdots x^{k_n}(\partial_{k_1}\cdots\partial_{k_n}R_{0k0l})(g), \\
    \label{eqn:FullResult0iComponent}
    h_{0i}&=-2\sum_{n=0}^\infty\frac{n+2}{(n+3)!}x^kx^lx^{k_1}\cdots x^{k_n}(\partial_{k_1}\cdots\partial_{k_n}R_{0kil})(g), \\
    \label{eqn:FullResultijComponent}
    h_{ij}&=-2\sum_{n=0}^\infty\frac{n+1}{(n+3)!}x^kx^lx^{k_1}\cdots x^{k_n}(\partial_{k_1}\cdots\partial_{k_n}R_{ikjl})(g).
\end{align}
We refer the reader to \cite{RAKH2014} for a more detailed discussion of FNC. In our case where the expected GW frequency is below the GHz-regime, we can apply the long wavelength approximation. That means, the Riemann tensor is independent of the spatial coordinates and the expansion can be cut off at second order. This leads to a vastly simplified metric (eqn. \ref{eqn:PDFMetric}), which can be used for calculating the Gertsenshtein current. Note, however, that the full expansion is needed in the GHz-regime and above. More details can be found in \cite{BERLIN2023, BERLIN2021}.

\section{Appendix B: Cavity Perturbation Theory}
\label{app:CavityPerturbationTheory}
When a GW propagates through a cavity, it changes the boundary conditions of the electromagnetic field. The eigenmodes of the deformed cavity are in general different from the eigenmodes of the unperturbed one. However, the GW strains are very small ($\lesssim\mathcal{O}(10^{-21})$), so cavity perturbation theory can be applied. That means, we can express the perturbed modes as series expansions of the unperturbed modes. We are then interested in the resulting overlap given by the coefficients of the expansion. An important result of this procedure is that the perturbed mode $\vb{E}'_n$ appears to be strongly coupled to its unperturbed counterpart $\vb{E}_n$, but also has contributions from other modes $\vb{E}_m$ with $m\neq n$. \\
There are several approaches to construct such an expansion. We will use the method given in \cite{GOUBAU1961} as it is consistent with the method applied in \cite{BERN2002}. The main idea is to find an expression for the deformed boundary conditions at the position of the unperturbed shell. The advantage of this approach is that we do not have to deal with a perturbed volume $V'_{\text{cav}}$ and can therefore work with $V_{\text{cav}}$ throughout the calculation. \\
We discuss the formalism in detail here, since \cite{GOUBAU1961} contains some inconsistencies. We further present the arguments in a new and improved way using a modern notation.

\subsection{The Perturbed Boundary Condition}
The unperturbed shell has surface $S$ while $S'$ denotes the surface of the perturbed shell. We note that the electromagnetic field in both cavities is described by the boundary value problem (BVP)
\begin{align}
    \label{eqn:PerturbedandUnperturbedBVP}
    \nabla\times\vb{E}_n&=ck_n\vb{B}_n, \qquad & \qquad \nabla\times\vb{E}'_n&=ck'_n\vb{B}'_n,\notag\\
    \nabla\times\vb{B}_n&=\frac{k_n}{c}\vb{E}_n, \qquad & \qquad \nabla\times\vb{B}'_n&=\frac{k'_n}{c}\vb{E}'_n, \\
    \vb{n}\times\vb{E}_n\vert_S&=0, \qquad & \qquad \vb{n}'\times\vb{E}'_n\vert_{S'}&=0. \notag 
\end{align}
We will use $\omega_n=ck_n$ instead of $k_n$ from now on. Our goal is to find the equivalent of the boundary condition $\vb{n}'\times\vb{E}'_n\vert_{S'}=0$ on $S$. Since $S$ is supposed to be a (at least piecewise) smooth manifold, we can parameterise it with two variables $\lambda_1$ and $\lambda_2$. We then define two differentiable curves
\begin{align*}
    \vb{u}_1 &:= \vb{u}_{\lambda_2}(\lambda_1):=\vb{S}(\lambda_1,\lambda_2)\vert_{\lambda_2\text{ fixed}}, \\
    \vb{u}_2 &:= \vb{u}_{\lambda_1}(\lambda_2):=\vb{S}(\lambda_1,\lambda_2)\vert_{\lambda_1\text{ fixed}},
\end{align*}
such that the tangential vectors
\begin{figure}[b!]
    \begin{center}
        \includegraphics[width=0.9\textwidth]{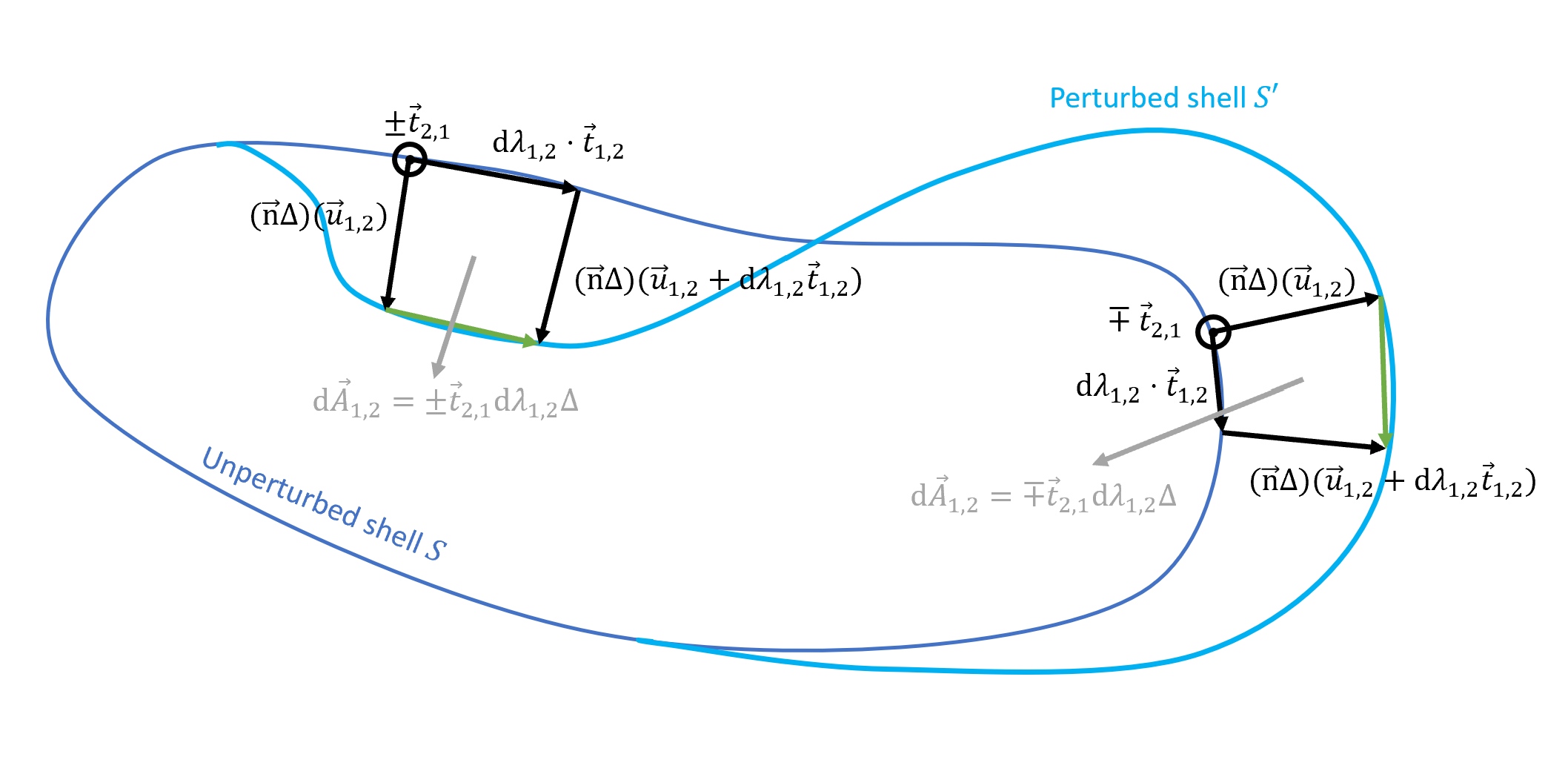}
        \caption{\textit{Construction of the boundary conditions for the perturbed mode. The idea is to use the area elements $\dd\vb{A}_{1,2}$, Stoke's theorem and that $\vb{E'_n}$ vanishes on the perturbed shell to find an expression for $\vb{n}\times\vb{E}'_n\vert_S$ on the unperturbed shell. This sketch should help to track the correct signs for inward/outward deformation and $\vb{t}_1$/$\vb{t}_2$-direction respectively.}}
        \label{fig:BoundaryConditions}
    \end{center}
\end{figure}
\begin{equation*}
    \vb{t}_1:=\firstdevp{\vb{u}_1}{\lambda_1}, \qquad \qquad \vb{t}_2=\firstdevp{\vb{u}_2}{\lambda_2}
\end{equation*}
define a right-handed orthonormal system ($\vb{t}_1,\vb{t}_2,\vb{n}$), where $\vb{n}$ is the surface normal. The displacement is described by $\Delta(\vb{x})$, which gives the absolute value of the shell deformation at a point $\vb{x}$ on the surface. We set $\Delta<0$ for inward and $\Delta>0$ for outward deformations. Throughout the following discussion, we will assume that $|\Delta(\vb{x})|\ll 1$. \\
We start by going the infinitesimal distances
\begin{equation*}
    \dd\vb{u}_1 = \vb{t}_1\dd\lambda_1, \qquad \qquad \qquad \dd\vb{u}_2 = \vb{t}_2\dd\lambda_2
\end{equation*}
on the unperturbed surface $S$. We then move along a closed path using $\vb{n}\Delta$ to jump on the perturbed surface $S'$. In figure \ref{fig:BoundaryConditions}, it is shown for inward and outward deformation. We consider the surface elements within the path, which are given by
\begin{equation*}
    \dd\vb{A}_1=\pm\vb{t}_2\dd \lambda_1\Delta, \qquad \qquad \dd\vb{A}_2=\mp\vb{t}_1\dd \lambda_2 \Delta.
\end{equation*}
Note that the upper sign corresponds to the inward direction and the lower sign to the outward direction. The key idea now is to apply Stokes theorem. It is useful to look at fig. \ref{fig:BoundaryConditions} to track the signs correctly. Weighting the surface elements $\dd\vb{A}_1$ and $\dd\vb{A}_2$ with $\nabla\times\vb{E}'_n$ leads to
\begin{align*}    
    \nabla\times\vb{E}'_n\cdot\dd\vb{A}_1&= \pm\nabla\times\vb{E}'_n\cdot\vb{t}_2\Delta\dd \lambda_1 \\
    &= \pm\vb{E}'_n\Delta\vb{n}\mp\vb{E}'_n\Delta\vb{n} \mp \dd\lambda_1\firstdevp{ }{\lambda_1}(\vb{E}'_n\vb{n}\Delta)\mp\vb{E}'_n\vb{t}_1\dd \lambda_1 \\
    &= \mp\vb{E}'_n\vb{t}_1\dd \lambda_1 \mp \dd\lambda_1\firstdevp{ }{\lambda_1}(\vb{E}'_n\vb{n}\Delta), \\                
   \nabla\times\vb{E}'_n\cdot\dd\vb{A}_2&= \mp\nabla\times\vb{E}'_n\cdot\vb{t}_1\Delta\dd\lambda_2 \\
    &= \pm\vb{E}'_n\Delta\vb{n}\mp\vb{E}'_n\Delta\vb{n} \mp \dd\lambda_2\firstdevp{ }{\lambda_2}(\vb{E}'_n\vb{n}\Delta)\mp\vb{E}'_n\vb{t}_2\dd \lambda_2 \\
    &= \mp\vb{E}'_n\vb{t}_2\dd \lambda_2 \mp \dd \lambda_2\firstdevp{ }{\lambda_2}(\vb{E}'_n\vb{n}\Delta),
\end{align*}
where we expanded $(\vb{E}'_n\vb{n}\Delta)(\vb{u}_{1,2}+\dd\lambda_{1,2}\vb{t}_{1,2})$ up to first order and used that $\vb{E}'_n\vert_S'=0$ on the perturbed surface (see fig. \ref{fig:BoundaryConditions}).
By eliminating $\dd \lambda_1$ and $\dd\lambda_2$, we find
\begin{align*}
    \vb{E}'_n\cdot\vb{t}_1&=-\nabla\times\vb{E}'_n\cdot\vb{t}_2\Delta-\firstdevp{ }{\lambda_1}(\vb{E}'_n\vb{n}\Delta), \\
    \vb{E}'_n\cdot\vb{t}_2&=\nabla\times\vb{E}'_n\cdot\vb{t}_1\Delta-\firstdevp{ }{\lambda_2}(\vb{E}'_n\vb{n}\Delta).
\end{align*}
These results can be now combined to
\begin{align*}
    \vb{E}'_n&=(\vb{E}'_n\vb{t}_1)\cdot\vb{t}_1+(\vb{E}'_n\vb{t}_2)\cdot\vb{t}_2+(\vb{E}'_n\vb{n})\cdot\vb{n} \\
    &= -(\nabla\times\vb{E}'_n\vb{t}_2\Delta)\cdot\vb{t}_1+(\nabla\times\vb{E}'_n\vb{t}_1\Delta)\cdot\vb{t}_2 \\
    &\qquad\qquad -\firstdevp{ }{\lambda_1}(\vb{E}'_n\vb{n}\Delta)\cdot\vb{t}_1-\firstdevp{ }{\lambda_2}(\vb{E}'_n\vb{n}\Delta)\cdot\vb{t}_2+(\vb{E}'_n\vb{n})\cdot\vb{n}.
\end{align*}
On the shell of the unperturbed cavity, this expression reads
\begin{equation*}
    \vb{E}'_n\vert_S=\vb{n}\times(\nabla\times\vb{E}'_n)\Delta\vert_S-\nabla(\vb{E}'_n\vb{n}\Delta)\vert_S+(\vb{E}'_n\vb{n})\cdot\vb{n}\vert_S,
\end{equation*}
where we used the standard gradient in the coordinate system ($\vb{t}_1$, $\vb{t}_2$, $\vb{n}$) together with the identity $\vb{a}\times(\vb{b}\times\vb{c})=\vb{b}\cdot(\vb{a}\cdot\vb{c})-\vb{c}\cdot(\vb{a}\cdot\vb{b})$. Inserting eqn.  \ref{eqn:PerturbedandUnperturbedBVP} finally yields the perturbed version of the boundary condition $\vb{n}\times\vb{E}_n\vert_S=0$. The result is 
\begin{equation*}           
   \vb{n}\times\vb{E}'_n\vert_S= \Delta(\omega_n\vb{B}_n\times\vb{n})\times\vb{n}\vert_S+\nabla(\vb{E}_n\vb{n}\Delta)\times\vb{n}\vert_S,
\end{equation*}
so the perturbed electric field does not vanish in the unperturbed shell. This will now help us to find a series expansion for $\vb{E}'_n$ (or $\vb{B}'_n$) in terms of $\vb{E}_n$ (or $\vb{B}_n$). Note that we have dropped the primes on the right hand side as we assumed $\Delta$ to be small. It is therefore sufficient to consider leading order terms only.

\subsection{Solving the Boundary Value Problem}
According to the general idea of perturbation theory, we can decompose the perturbed eigenmodes as \cite{BERN2002}
\begin{align*}
    \vb{E}'_n &= \vb{E}_n+\cormode{\vb{E}}+\mathcal{O}(\sigma^2), \\
    \vb{B}'_n &= \vb{B}_n+\cormode{\vb{B}}+\mathcal{O}(\sigma^2), \\
    \omega'_n &= \omega_n+\cormode{\omega}+\mathcal{O}(\sigma^2).
\end{align*}
Substituting this into the perturbed BVP and using the unperturbed BVP (see eqn. \ref{eqn:PerturbedandUnperturbedBVP}), we obtain a BVP for the first order corrections $\cormode{\vb{E}}$ and $\cormode{\vb{B}}$. It can be written as 
\begin{align}
    \label{eqn:PerturbedDE1}
    \nabla\times\cormode{\vb{E}}-\omega_n\cormode{\vb{B}}&=\cormode{\omega}\vb{B}_n, \\
    \label{eqn:PerturbedDE2}
    \nabla\times\cormode{\vb{B}}-\frac{\omega_n}{c^2}\cormode{\vb{E}}&=\frac{\cormode{\omega}}{c^2}\vb{E}_n, \\
    \label{eqn:PerturbedBC}
    \vb{n}\times\cormode{\vb{E}}\vert_S=\omega_n\vb{V_n}\vert_S,
\end{align}
where we again consider leading order terms in $\Delta$ and $\sigma$ only. In order to abbreviate notation, we have defined 
\begin{equation}
    \label{eqn:BoundaryAbbreviation}
    \vb{V}_n:=\Delta (\vb{B}_n\times\vb{n})\times\vb{n}\vert_S+\frac{1}{\omega_n}\nabla(\vb{E}_n\vb{n}\Delta)\times\vb{n}\vert_S
\end{equation}
here. We can now expand the first order corrections in terms of the unperturbed modes, i.e. 
\begin{align}
    \label{eqn:EExpansion}
    \cormode{\vb{E}}&=\sum_m\alpha_{nm}\vb{E}_m, \\
    \label{eqn:BExpansion}
    \cormode{\vb{B}}&=\sum_m\beta_{nm}\vb{B}_m, \\
    \label{eqn:kExpansion}
    \cormode{\omega}&=\sum_m\kappa_{nm}\omega_m.
\end{align}
The remaining task then is to find the coefficients $\alpha_{nm}$, $\beta_{nm}$ and $\kappa_{nm}$.  We start by integrating equation \ref{eqn:PerturbedDE1} over $\vb{B}_m$ such that
\begin{align}
    \label{eqn:PerturbedPrelResult1}
    \intcav\dd^3x\vb{B}_m\cdot\nabla\times\cormode{\vb{E}}-\omega_n\intcav\dd^3x\vb{B}_m,\cdot\cormode{\vb{B}} &= \cormode{\omega}\delta_{nm}\intcav\dd^3x\vb{B}^2_n.
\end{align}
Equivalently, we can integrate equation \ref{eqn:PerturbedDE2} over $\vb{E}_m$ which leads to a similar expression with $\vb{B}$ and $\vb{E}$ exchanged. 
Using standard nabla identities and Gauss's law, we can rewrite the first integral of eqn. \ref{eqn:PerturbedPrelResult1} as
\begin{align*}
    \intcav\dd^3x\vb{B}_m\cdot\nabla\times\cormode{\vb{E}}&=-\intcavbound\dd\vb{S}(\vb{B}_m\times\cormode{\vb{E}})+\intcav\dd^3x\cormode{\vb{E}}\nabla\times\vb{B}_m. 
\end{align*}
To evaluate the surface integral, we can use the boundary conditions in \ref{eqn:PerturbedandUnperturbedBVP} and \ref{eqn:PerturbedBC}. Note that there is now a difference between the E-field and B-field because
\begin{align*}
    \dd\vb{S}(\vb{B}_m\times\cormode{\vb{E}})&=\vb{n}\cdot(\vb{B}_m\times\cormode{\vb{E}})\dd S = \vb{B}_m\cdot(\cormode{\vb{E}}\times\vb{n})\dd S = -\omega_n\vb{B}_m\cdot\vb{V}_n\dd S \\
    \dd\vb{S}(\vb{E}_m\times\cormode{\vb{B}})&=\vb{n}\cdot(\vb{E}_m\times\cormode{\vb{B}})\dd S=\cormode{\vb{B}}\cdot(\vb{n}\times\vb{E}_m)\dd S = 0.
\end{align*}
With these results and using eqn. \ref{eqn:PerturbedandUnperturbedBVP}, we can write equation \ref{eqn:PerturbedPrelResult1} as
\begin{align*}
    \frac{\omega_m}{c^2}\intcav\dd^3x\cormode{\vb{E}}\vb{E}_m&- \omega_n\intcav\dd^3x\cormode{\vb{B}}\cdot\vb{B}_m \\ &= \cormode{\omega}\delta_{nm}\intcav\dd^3x\vb{B}^2_n-\omega_n\intcavbound\dd S \vb{B}_m\cdot\vb{V}_n, \\
    \omega_m\intcav\dd^3x\cormode{\vb{B}}\vb{B}_m&-\frac{\omega_n}{c^2}\intcav\dd^3x\cormode{\vb{E}}\vb{E}_m \\ &= \frac{\cormode{\omega}}{c^2}\delta_{nm}\intcav\dd^3x\vb{E}^2_n,
\end{align*}
where we also gave the corresponding expression for the B-field. The next step is to insert the expansions eqn. \ref{eqn:EExpansion}-\ref{eqn:BExpansion}. We can use eqn. \ref{eqn:Normalization} to simplify the notation and arrive at
\begin{align}
    \label{eqn:DefiningEquation1}
    \frac{\omega_m}{c^2}\alpha_{nm}\frac{2U_m}{\epsilon_0}-\omega_n\beta_{nm}2\mu_0U_m &= \delta_{nm}\cormode{\omega}2\mu_0U_m+\frac{2U_m}{\epsilon_0}\frac{\omega_n}{c^2}\mathcal{C}_{nm}, \\
    \label{eqn:DefiningEquation2}
    \omega_m\beta_{nm}2\mu_0U_m-\frac{\omega_n}{c^2}\alpha_{nm}\frac{2U_m}{\epsilon_0}&=\frac{\cormode{\omega}}{c^2}\delta_{nm}\frac{2U_n}{\epsilon_0},
\end{align}
where a new coupling coefficient is defined by
\begin{equation}
    \label{eqn:OriginalCouplingCoefficient}
    \mathcal{C}_{nm}:=-\frac{c^2}{2U_m}\intcavbound\dd S\epsilon_0\vb{B}_m\vb{V}_n.
\end{equation}
To find the coefficients $\alpha_{nm}$, $\beta_{nm}$ and $\kappa_{nm}$, we have to solve eqn. \ref{eqn:DefiningEquation1} and \ref{eqn:DefiningEquation2}. Therefore, we have to distinguish between the cases $n=m$ and $n\neq m$. We start with the latter, which yields
\begin{align}
    \alpha_{nm}&=\frac{\omega_m \omega_n}{\omega^2_m-\omega^2_n}\mathcal{C}_{nm}, \\ \label{eqn:BetaCoefficients}\beta_{nm}&=\frac{\omega^2_n}{\omega^2_m-\omega^2_n}\mathcal{C}_{nm}.
\end{align}
The case $n=m$ needs a bit more work. From eqn. \ref{eqn:DefiningEquation1} and \ref{eqn:DefiningEquation2}, we directly find
\begin{equation*}
    \cormode{\omega}=-\frac{1}{2}\omega_n\mathcal{C}_{nn}.
\end{equation*}
This leads to a solution for $\kappa_{nm}$ and an expression for $\alpha_{nn}$ and $\beta_{nn}$, which read
\begin{align*}
    \kappa_{nm}&=-\frac{1}{2}\delta_{nm}\mathcal{C}_{nm}, \\
    \alpha_{nn}&=\beta_{nn}+\frac{1}{2}\mathcal{C}_{nn}.
\end{align*}
However, we have to fix another degree of freedom to get a final result for the remaining coefficients. That is because we have not yet chosen a normalization for the perturbed fields. An appropriate choice is to define
\begin{equation}
    \label{eqn:PerturbedNormalization}
    \intcav\dd^3x\vb{E}'^2_n:=\frac{2U_n}{\epsilon_0}=\intcav\dd^3x\vb{E}^2_n.
\end{equation}
By observing that
\begin{equation*}
    \frac{2U_n}{\epsilon_0}=\intcav\dd^3x(\vb{E}_n+\cormode{\vb{E}})^2 = (1+2\alpha_{nn})\frac{2U_n}{\epsilon_0},
\end{equation*}
we find that the diagonal coefficients are asymmetric and given by
\begin{equation}
    \label{eqn:CoefficientsDiagonalComponents}
    \alpha_{nn}=0, \qquad \qquad \qquad \beta_{nn}=-\frac{1}{2}\mathcal{C}_{nn}.
\end{equation}
Finally, we write the perturbed solutions in terms of the time modes $e_n(t)$ and $b_n(t)$. By substituting the expansions \ref{eqn:EExpansion}-\ref{eqn:kExpansion} into eqn. \ref{eqn:TimeExpansion}, we end up with
\begin{align}
    \label{eqn:PerturbedTimeModee}
    e'_n(t)&=e_n(t)+\sum_{m\neq n}\alpha_{nm}\frac{U_m}{U_n}e_m(t),  & \qquad  \alpha_{nm}&=\frac{\omega_n \omega_m}{\omega^2_m-\omega^2_n}\mathcal{C}_{nm}, \\
     \label{eqn:PerturbedTimeModeb}
     b'_n(t)&=b_n(t)-\frac{1}{2}\mathcal{C}_{nn}b_n(t)+\sum_{m\neq n}\frac{U_m}{U_n}\beta_{nm}b_m(t), & \qquad  \beta_{nm}&=\frac{\omega^2_n}{\omega^2_m-\omega^2_n}\mathcal{C}_{nm}, \\
    \omega'_n&=\omega_n-\frac{1}{2}\omega_n\mathcal{C}_{nn}.
\end{align}
The remaining task is to determine the connection coefficients $\mathcal{C}_{nm}$.

\subsection{The Connection Coefficient}
We return to eqn. \ref{eqn:BoundaryAbbreviation} and eqn. \ref{eqn:OriginalCouplingCoefficient}, which define the connection coefficient $\mathcal{C}_{nm}$. The full expression reads
\begin{equation*}
    \mathcal{C}_{nm}=-\frac{c^2}{2U_m}\intcavbound\dd S\epsilon_0\vb{B}_m\Big[ (\Delta\vb{B}_n\times\vb{n})\times\vb{n}+\frac{1}{\omega_n}\nabla(\vb{n}\vb{E}_n \Delta)\times\vb{n} \Big].
\end{equation*}
We can write this in a shorter form by using the boundary condition $\vb{B}_n\cdot\vb{n}\vert_S=0$ for the unperturbed cavity. The left integral can be then written as
\begin{equation*}
    \intcavbound\dd S\cdot\vb{B}_m(\vb{B}_n\times\vb{n})\times\vb{n}\Delta = -\intcavbound\dd S\cdot\vb{B}_m\vb{B}_n\Delta.
\end{equation*}
For the right integral, we use that $\dd\vb{S}=\vb{n}\dd S$ and eqn. \ref{eqn:PerturbedandUnperturbedBVP} to find
\begin{align*}
    \intcavbound\dd S&\cdot\epsilon_0\vb{B}_m\nabla(\vb{n}\vb{E}_n\Delta)\times\vb{n} \\
    &= \omega_m\intcavbound\dd S\cdot\frac{\epsilon_0}{c^2}\Delta(\vb{n}\vb{E}_n)(\vb{n}\vb{E}_m)-\intcavbound\dd\vb{S}\cdot\nabla\times((\vb{n}\vb{E}_n)\vb{B}_m\Delta).
\end{align*}
Using the boundary condition $\vb{E}_{n,m}\times\vb{n}\vert_S=0$, we can write 
\begin{equation*}
    (\vb{n}\cdot\vb{E}_n)(\vb{n}\cdot\vb{E}_m)\vert_S=\vb{E}_n\cdot\vb{E}_m\vert_S.
\end{equation*}
in the first integral. The second integral vanishes due to Stoke's law. Combining all results leads to the relation
\begin{equation*}
    \mathcal{C}_{nm}=\frac{1}{2U_m}\intcavbound\dd S\cdot\Delta\Big[ \frac{1}{\mu_0}\vb{B}_n\vb{B}_m-\frac{\omega_m}{\omega_n}\epsilon_0\vb{E}_n\vb{E}_m \Big].
\end{equation*}
Note that in cases where $\omega_m\approx \omega_n$ like in heterodyne cavity experiments, we can write the simplified form 
\begin{equation}    
    \label{eqn:GeneralConnectionCoefficient}
    \mathcal{C}_{nm}\approx\frac{1}{2U_m}\intcavbound\dd S\cdot\Delta\Big[ \frac{1}{\mu_0}\vb{B}_n\vb{B}_m-\epsilon_0\vb{E}_n\vb{E}_m \Big].
\end{equation}

\end{document}